# Cooperative STT and SOT switching in perpendicular magnetic tunnel junctions: Role of the pulse-end magnetization state

Zeyi Zhu[1], Ding Ma[1], Kai Zhang[1], Mingwei Wu[1], Yi Huang[1], Jiarui Niu[1], Shuai Zhang[1], Mengdi Yin[1], Ka-San Mak[1], Lina Chen[2,3], Bo Liu[4], Tiejun Zhou[4], Yongbing Xu[5], and Ronghua Liu[1,5,*]

[1]State Key Laboratory of Spintronics and School of Physics and Collaborative Innovation Center of Advanced Microstructures, Nanjing University, Nanjing 210093, China

[2]School of Science, Nanjing University of Posts and Telecommunications, Nanjing 210023, China

[3]National Laboratory of Solid State Microstructures, Nanjing University, Nanjing 210093, China

[4]State Key Laboratory for Spintronic Devices and Technologies, Hangzhou 311305, China

[5]State Key Laboratory of Spintronics, Nanjing University, Suzhou 215163, China

*Email: rhliu@nju.edu.cn

Abstract

Spin–orbit torque MRAM (SOT-MRAM) is a leading candidate for next-generation nonvolatile memory, offering high speed, endurance, and architectural compatibility. However, in conventional SOT switching, the magnetization remains near the in-plane region at pulse termination, making the final state highly sensitive to post-pulse relaxation dynamics and prone to back-switching. To overcome this, we propose a field-free scheme in which the transverse SOT drives large-angle precessional excitation while the perpendicular spin-transfer torque (STT) biases the trajectory toward the reversed −z state. Micromagnetic simulations reveal a nonlinear switching boundary in the $J_{STT}$–$J_{SOT}$ parameter space, originating from the distinct dynamical roles of the two torques: SOT primarily governs the excitation and crossing of the dynamical separatrix, whereas STT controls the terminal trajectory and final-state selection. An analytical macrospin model, based on the stability analysis of the current-induced equilibrium, reproduces the critical-boundary trends as functions of current density, Gilbert damping α, and uniaxial anisotropy $K_u$, and distinguishes dynamic anti-damping and static instability branches. Systematic analyses of pulse duration, damping, anisotropy, and the STT–SOT balance further demonstrate that reliable ultrafast switching requires not only sufficient excitation to cross the separatrix before pulse termination, but also precise control of the pulse-end magnetization state to minimize post-pulse relaxation. These results establish that the pulse-end state, rather than the instantaneous torque amplitude, is the decisive factor governing switching speed and reliability in coupled STT–SOT systems.

## I. Introduction

Magnetoresistive random-access memory (MRAM) is a prominent candidate for next-generation nonvolatile memory, owing to its exceptional endurance, fast write/read speeds, and seamless compatibility with advanced semiconductor architectures [1-4]. Among various MRAM technologies, the three-terminal SOT-MRAM stands out for its high-speed writing capability, high endurance, and favorable read/write margins. This is achieved by injecting an in-plane charge current through a heavy-metal underlayer, which generates a spin current via the spin Hall effect to exert a torque on the adjacent free magnetic layer [5-10]. Crucially, this geometry separates the read and write paths, significantly reducing electrical stress across the sensitive magnetic tunnel junction (MTJ) barrier. These attributes make SOT-MRAM particularly attractive for demanding applications such as last-level processor caches and embedded nonvolatile memory [11-14].

However, a fundamental challenge persists in perpendicular magnetic anisotropy (PMA) systems: in the conventional SOT geometry illustrated in Fig. 1(a), an external in-plane magnetic field $H_x$ or an effective field is typically required to break the symmetry and enable deterministic magnetization switching via SOT alone [15-18]. Although various field-free schemes have been proposed [11,19-21], the magnetization switching process remains heavily influenced- if not determined- by the post-pulse relaxation dynamics, which are intimately tied to this effective in-plane field and the magnetization equilibrium position at the moment the current pulse terminates. Even when initial reversal is successful, the magnetization vector often resides near the in-plane region at the pulse end. Following pulse removal, precession in the presence of $H_x$ can drive the magnetization back across the film plane, ultimately returning it to its initial state [22,23]. As depicted in Fig. 1(b), this back-switching behavior occurs at higher current magnitudes and severely constrains the effective write-current window. Figure 1(c) illustrates a representative trajectory of this process, with the color evolving from blue to red as time progresses. The final state is determined by a post-pulse relaxation trajectory characterized by oscillatory *z*-components induced by the in-plane effective field. Beyond the in-plane effective field, this relaxation phase is governed by a highly nonlinear dependence on parameters such as the Gilbert damping constant, the PMA constant, and the precise torque balance, giving rise to significant statistical variability. This "pulse-end-state-dependent" relaxation can narrow the effective write-current window and is intrinsically linked to undesirable back-switching events at high current densities, ultimately compromising write reliability [24]. Consequently, achieving robust switching demands precise control over the terminal stage of the magnetization trajectory- an objective that can be pursued through pulse-shape engineering, material optimization, or the introduction of an additional torque mechanism [25,26].

Previous studies have demonstrated that combining SOT with spin-transfer torque (STT) in perpendicular MTJs can enable field-free switching, reduce the SOT switching threshold, and improve sub-nanosecond switching dynamics [27-29]. However, the role of the pulse-end magnetization state and the subsequent relaxation trajectory in determining the effective write-current window remains insufficiently clarified. In this work, we propose and systematically investigate a field-free writing scheme that synergistically combines STT and SOT to address this challenge. We employ an analytical model and micromagnetic simulations to study the switching behavior in PMA-MTJs driven by this combined STT-SOT action. Our simulations reveal a complementary, cooperative mechanism: the SOT component efficiently provides the transverse torque to drive the free-layer magnetization from its initial perpendicular state into a near-in-plane state. Concurrently, the out-of-plane STT component biases the dynamical trajectory toward the target reversed state (±z), thereby assisting final-state selection and stabilizing the outcome. Building on this mechanistic insight, we analyze how key parameters—including current pulse duration, Gilbert damping constant, anisotropy constant, and the critical balance between STT and SOT current densities—govern switching behavior. This analysis elucidates how the switching boundary, the magnetization state at pulse end, and the subsequent relaxation process collectively determine the reversal efficiency and speed. Our results show that fast, reliable switching requires not only sufficient torque excitation (JSOT and JSTT) to drive the magnetization across the dynamical separatrix during the pulse, but also strategic control of the pulse-end trajectory to minimize terminal relaxation time after the pulse. These findings provide concrete guidelines for designing optimized STT-SOT-assisted MRAM devices with enhanced write margin, accelerated switching speed, and improved operational reliability.

## II. Micromagnetic Modeling and Simulation

Micromagnetic simulations were performed using OOMMF by solving the Landau–Lifshitz–Gilbert (LLG) equation including STT and SOT terms [30-32],

$$\frac{d\boldsymbol{m}}{dt} = -\gamma\mu_0\boldsymbol{m}\times\boldsymbol{H}_{\text{eff}} + \alpha\boldsymbol{m}\times\frac{d\boldsymbol{m}}{dt} - \gamma\mu_0 H_{\text{STT}}\boldsymbol{m}\times(\boldsymbol{m}\times\boldsymbol{p}) + \gamma\mu_0 H_{\text{SOT}}\boldsymbol{m}\times(\boldsymbol{m}\times\boldsymbol{\sigma}), \quad (1)$$

where $\boldsymbol{m}$ is the unit magnetization vector, $\gamma$ is the gyromagnetic ratio, $\mu_0$ is the vacuum permeability, $\alpha$ is the Gilbert damping constant, $\boldsymbol{H}_{\text{eff}}$ is the effective magnetic field, $\boldsymbol{P}$ is the STT spin-polarization direction, and $\boldsymbol{\sigma}$ is the spin-polarization direction associated with the damping-like SOT.

$$\mu_0 H_{\text{STT}} = \frac{\hbar\eta J_{\text{STT}}}{2eM_s t_F}, \quad \mu_0 H_{\text{SOT}} = \frac{\hbar\xi J_{\text{SOT}}}{2eM_s t_F} \quad (2)$$

where $\eta$ and $\xi$ denote the STT polarization efficiency and the damping-like SOT efficiency, respectively; $M_s$ is the saturation magnetization, and $t_F$ is the free-layer thickness. The magnetically active region was the CoFeB free layer, modeled as a circular disk with a diameter of 15 nm and a thickness of 1 nm. The free layer was discretized using a Cartesian finite-difference mesh with a cell size of $1 \times 1 \times 1$ nm$^3$. This cell size is smaller than the exchange length [33]

$$l_{\text{ex}} = \sqrt{2A_{\text{ex}} / \mu_0 M_s^2}, \quad (3)$$

where $A_{\text{ex}}$ is the exchange stiffness constant, and $M_s$ is the saturation magnetization. The $l_{\text{ex}}$ is estimated to be approximately 3.88 nm for the adopted material parameters, and is therefore sufficient to resolve the magnetization texture.

As illustrated schematically in Fig. 1(d), the physical device consists of a heavy-metal underlayer, a free ferromagnetic layer, a nonmagnetic spacer, and a fixed ferromagnetic layer. However, in the current OOMMF implementation, we explicitly discretized and dynamically evolved only the free layer. We set the CoFeB free-layer material parameters to $M_s$ = 1.3 × 106 A/m, $A_{ex}$ = 1.6 × 10-11 J/m, α = 0.1, and $K_u$ = 1.35 × $10^6$ J/$m^3$, with the easy axis along the $z$ direction. No external magnetic field was applied.

The fixed layer and the heavy-metal underlayer were incorporated implicitly through effective spin-torque terms in the LLG equation. Specifically, the STT term was characterized by a fixed spin-polarization direction $\boldsymbol{P}$ = (0, 0, −1), corresponding to the magnetization orientation of the reference layer, with a polarization efficiency $\eta$ = 0.25. We introduced the SOT term phenomenologically with a damping-like spin-polarization direction σ = (0, 1, 0), corresponding to a charge current applied along the +x direction, and an effective spin Hall efficiency $\xi$ = 0.4. We limited the maximum time step to 1 ps to ensure numerical stability. During the simulations, we recorded both the spatial magnetization distribution and the spatially averaged magnetization components to analyze the free-layer switching behavior.

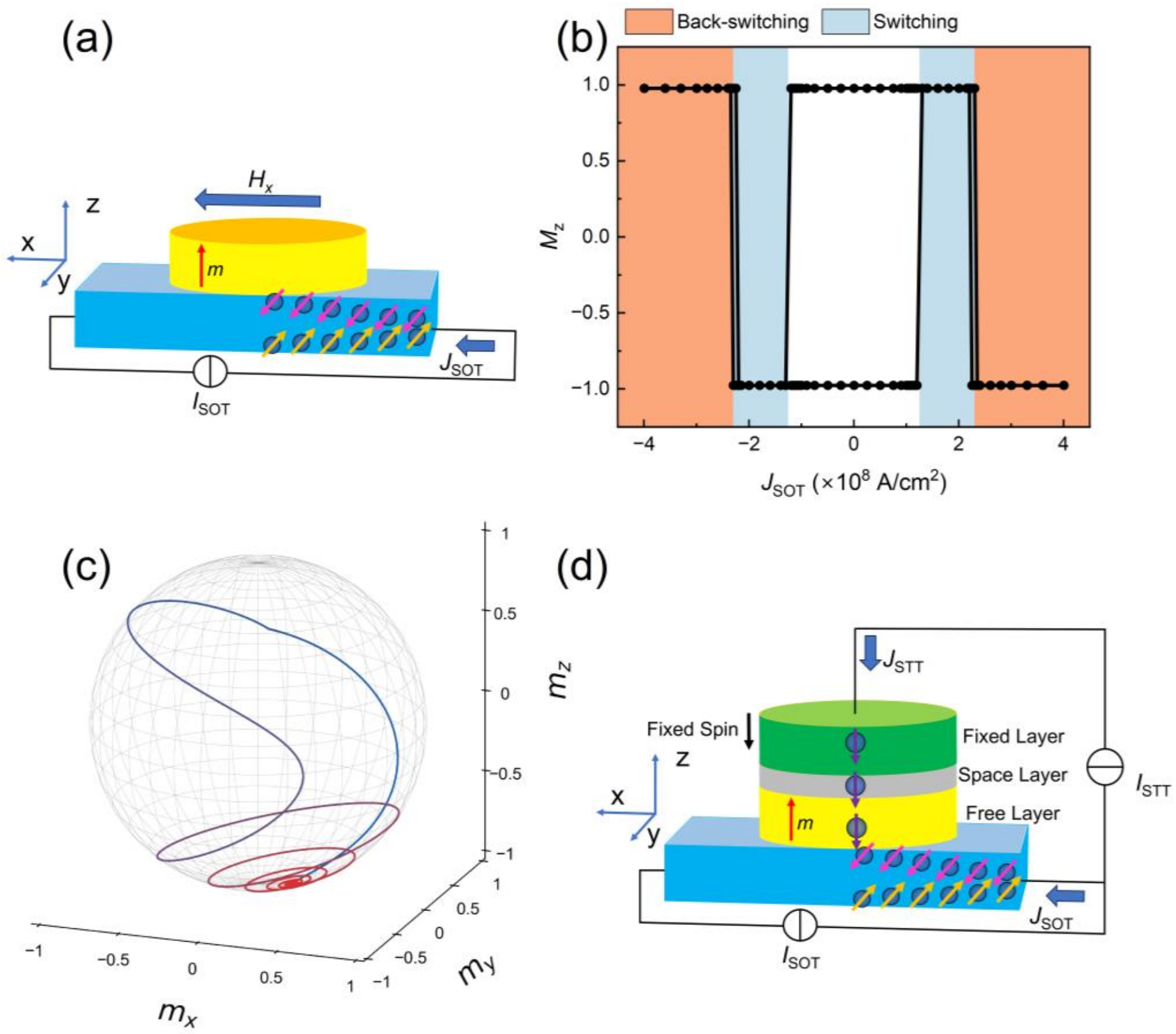

**Fig. 1 Comparison of conventional field-assisted SOT switching and field-free switching driven by combined STT and SOT.** (a) Schematic of the conventional SOT geometry, consisting of a perpendicular ferromagnetic layer on a heavy-metal channel. An in-plane current density $J_{SOT}$ generates SOT via the spin Hall effect, and an external in-plane magnetic field $H_x$ assists magnetization switching. (b) Final out-of-plane magnetization component as a function of $J_{SOT}$. The light-blue shaded regions indicate successful switching, whereas the orange shaded regions indicate back-switching at higher current magnitudes, where the magnetization returns to its initial state during post-pulse relaxation, limiting the switching window. (c) Representative magnetization trajectory on the unit sphere illustrating the back-switching process. After pulse removal, precession in the in-plane field $H_x$ drives the magnetization back across the in-plane region, followed by relaxation toward its initial state. The trajectory is colored chronologically from blue at the beginning to red at the end. (d) Schematic of the field-free STT-SOT device, comprising a heavy-metal channel, a perpendicular free layer, a nonmagnetic spacer, and a fixed layer polarized along –$z$. The current densities $J_{SOT}$ in the heavy-metal channel and $J_{STT}$ through the magnetic tunnel junction jointly drive magnetization switching without an external magnetic field.

## III. Results and Discussion

Conventional SOT switching in PMA systems faces fundamental symmetry constraints and typically requires an in-plane auxiliary field for deterministic reversal. This approach leaves the magnetization near the in-plane region at pulse termination, making the final state highly sensitive to post-pulse relaxation dynamics. The residual in-plane field can induce pronounced oscillations of the $z$-component during post-pulse relaxation, which may drive the magnetization back across the dynamical boundary and result in backward switching, an instability that becomes increasingly severe at higher current densities.

To overcome these limitations, we employ a field-free combined STT-SOT geometry, where the two torque components play distinct yet complementary roles. For a free layer with PMA, the magnetization initially in the $+z$ state must be driven out of its potential well and across the dynamical separatrix before the current pulse ends; otherwise, even large-amplitude transient precession may relax back to the initial state after the pulse is removed. In our scheme, the damping-like SOT, generated by the in-plane current via the spin Hall effect, efficiently provides transverse excitation and launches large-angle precession, while the perpendicular STT biases subsequent relaxation toward the reversed −z state and helps complete the switching process. The inset of Fig. 2 shows three-dimensional (3D) magnetization trajectories illustrating this cooperative mechanism.

### A. Switching phase diagram *vs.* critical current density

Figure 2 shows how the cooperative action of SOT and STT governs magnetization dynamics and thus determines the switching boundary in the two-current parameter space. For the current-density combination below the switching boundary [Fig. 2(a)], the magnetization of the MTJ free layer does not reverse under the 2-ns STT-SOT pulse. At 0.3 ns, upon pulse application, the in-plane SOT drives the magnetization toward the film plane. However, the combined STT-SOT effect is insufficient to overcome the effective field arising from the PMA. Consequently, the magnetization continues to precess along a direction slightly tilted relative to the $z$-axis. After the pulse is turned off, the magnetization returns to its initial state under the PMA field's restoring action, and no switching occurs. Correspondingly, the $z$-component of the SOT ($\tau_z^{\mathrm{SOT}}$), rapidly increases in magnitude upon pulse application, whereas that of the STT ($\tau_z^{\mathrm{STT}}$), remains comparatively small. Although $J_{\mathrm{STT}}$ and $J_{\mathrm{SOT}}$ remain constant during the pulse, both torque $z$-components vary continuously with the instantaneous orientation of $\boldsymbol{m}$ and gradually diminish as the magnetization settles into a small-angle precessional trajectory. After the pulse is removed, $\tau_z^{\mathrm{SOT}}$ and $\tau_z^{\mathrm{STT}}$ vanish, and the magnetization relaxes back to its initial $+z$ state under the PMA field. When $J_{\mathrm{SOT}}$ is increased to $7.2\times10^8$ A/cm$^2$ [Fig. 2(b)], the magnetization dynamics change markedly. The enhanced precession amplitude renders the combined torques strong enough to overcome the PMA field, leading to a rapid reversal of $m_z$ at about 0.8 ns. During this process, $m_x$ and $m_y$ exhibit significant transient variations, reflecting vigorous magnetization precession. Although the amplitudes of the applied $J_{\mathrm{STT}}$ and $J_{\mathrm{SOT}}$ are held constant, the resulting torques acting on $\boldsymbol{m}$ vary continuously with its instantaneous orientation. Crucially, the $z$-components of both SOT and STT increase as $\boldsymbol{m}$ tilts away from the $+z$ axis. This growth enables the negative $z$-component of the STT to eventually drive $\boldsymbol{m}$ sharply across the dynamical separatrix and into the basin of the $-z$ state. $\boldsymbol{m}$ does not become perfectly aligned with the −z direction while the current pulse is still on, because the STT and SOT persistently compete. After the pulse is removed, the current-induced torques vanish, and the magnetization relaxes to the stable $-z$ state, confirming successful switching.

A similar switching outcome can be achieved by increasing $J_{\mathrm{STT}}$ [Fig. 2(c)]. A key distinction lies in the switching time: reversal completes within 0.7 ns for the increased-$J_{\mathrm{STT}}$ case, faster than when increasing $J_{\mathrm{SOT}}$. Together with the evolution of $\tau_z^{\mathrm{SOT}}$ and $\tau_z^{\mathrm{STT}}$ shown in Figs. 2(a)-2(c), this temporal difference reveals the distinct dynamical roles of the two torques. The SOT predominantly drives $\boldsymbol{m}$ into large-angle precession toward the in-plane region, accompanied by pronounced oscillations of its out-of-plane component. Concurrently, the sustained negative $z$-component of the STT biases the precessional trajectory toward the $-z$ basin and actively facilitates magnetization reversal. The two torques therefore play complementary roles: SOT primarily promotes the large-angle excitation of $\boldsymbol{m}$, whereas STT steers the resulting trajectory toward the reversed state. This cooperative action explains the faster switching observed when $J_{\mathrm{STT}}$ increases.

To further elucidate the influence of STT and SOT current densities on magnetization reversal, Fig. 2 (d) plots a switching phase diagram in the $J_{STT}$-$J_{SOT}$ plane. The diagram confirms that STT and SOT compensate for each other in lowering the switching threshold. The blue and orange regions correspond to the no-switching and switching regimes, respectively. The solid line with circular markers denotes the switching boundary obtained from micromagnetic simulations, whereas the dashed line represents the boundary derived from the analytical model, which will be introduced in a later section. A higher SOT current reduces the critical STT current required for switching, and vice versa, demonstrating the cooperative action of the two torques in reducing the overall switching barrier. Notably, the switching boundary is distinctly nonlinear, indicating that the two torques cannot be described by a simple linear superposition of independent driving terms. Deterministic reversal occurs only if the combined pulse drives $\boldsymbol{m}$ onto the reversed side of the dynamical separatrix before the pulse ends. Importantly, this STT-SOT scheme shows no backward switching. This immunity arises because the absence of an in-plane auxiliary field eliminates the orientation-dependent $z$-component torque that, in conventional SOT-only switching, can drive $\boldsymbol{m}$ back toward the initial state during the relaxation phase.

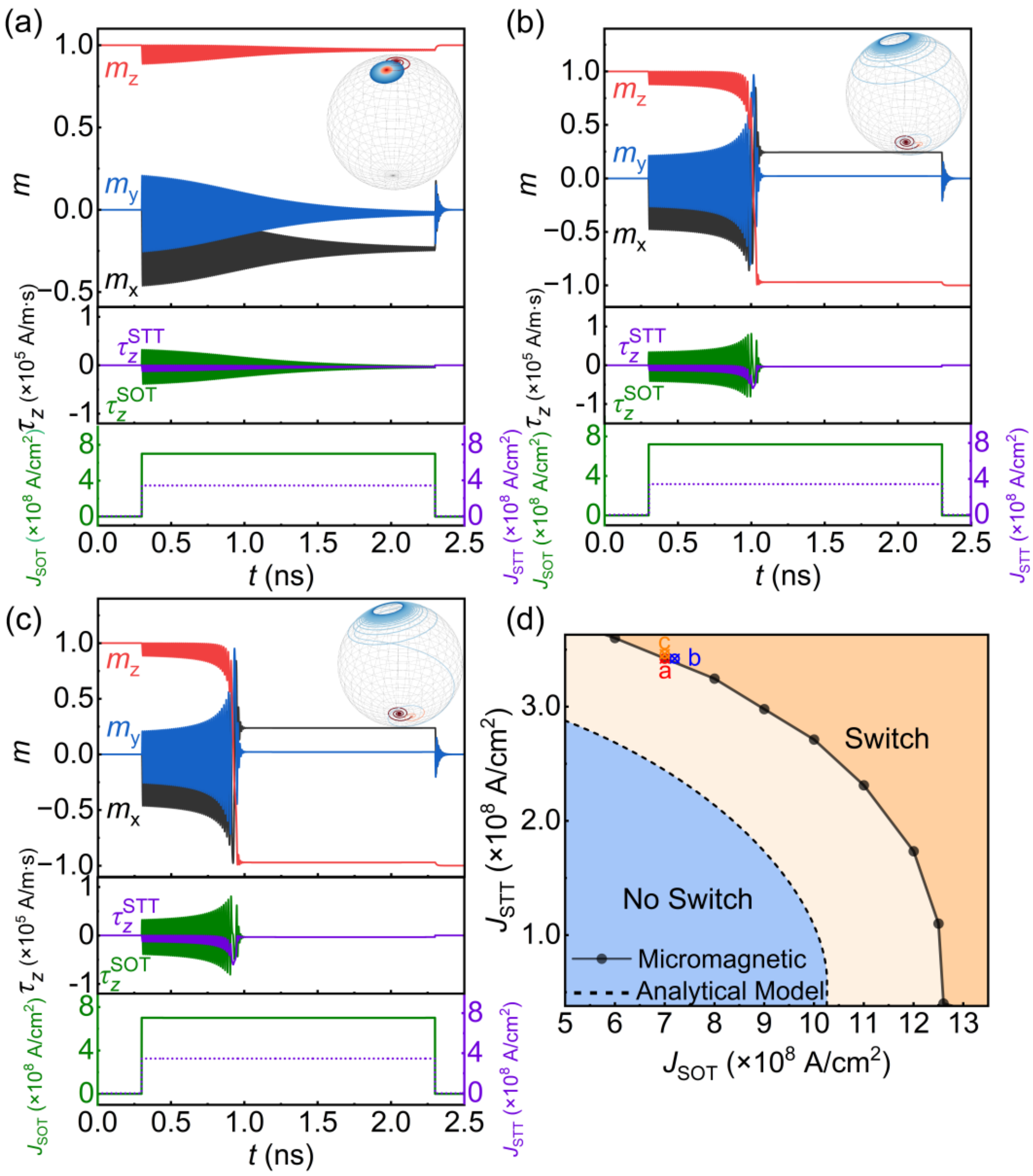

**Fig. 2 Magnetization switching dynamics and phase diagram under combined SOT and STT driving.** (a)-(c) Time evolution of the normalized magnetization components $m_x$, $m_y$, and $m_z$ (top), the $z$-components of the STT and SOT effective torques, $\tau_z^{SOT}$ and $\tau_z^{STT}$ (middle), and the applied 2-ns current-density pulses $J_{STT}$ and $J_{SOT}$ (bottom) for three representative current-density pairs: (a) $J_{STT} = 3.42 \times 10^8$ A/cm$^2$, $J_{SOT} = 7.0 \times 10^8$ A/cm$^2$; (b) $J_{STT} = 3.42 \times 10^8$ A/cm$^2$, $J_{SOT} = 7.2 \times 10^8$ A/cm$^2$, and (c) $J_{STT} = 3.47 \times 10^8$ A/cm$^2$, $J_{SOT} = 7.0 \times 10^8$ A/cm$^2$. Micromagnetic simulations used a damping constant $\alpha = 0.1$ and uniaxial anisotropy $K_u = 1.35 \times 10^6$ J/m$^3$. Insets show the corresponding 3D magnetization trajectories on the unit sphere, with the color gradient from blue (initial state) to red (final state) indicating time progression. (d) Switching phase diagram in the $J_{SOT}$-$J_{STT}$ parameter space. The dashed line represents the switching boundary derived from the analytical model assuming an infinite pulse duration, while the solid line with circular markers denotes the boundary obtained from micromagnetic simulations with 2 ns-duration pulses. The blue and orange shaded areas indicate the no-switching and switching regimes, respectively; the light-shaded band between the two boundaries highlights the parameter region where the thresholds differ slightly due to the finite (vs. infinite) pulse duration. Points **a**, **b**, and **c** mark the parameter sets used in panels (a)-(c), respectively.

### B. Switching phase diagram *vs.* current-pulse duration

In practical MRAM applications, the current-pulse duration is a critical parameter governing write speed, switching reliability, and energy consumption [25,34,35]. To elucidate its role, we investigated how pulse duration, JSTT, and JSOT affect magnetization-reversal dynamics. As illustrated in Fig. 3(a), for $J_{\mathrm{STT}} = 4 \times 10^8$ A/cm$^2$ and $J_{\mathrm{SOT}} = 5.4 \times 10^8$ A/cm$^2$ with a pulse duration of 0.23 ns, the combined current pulse excites significant magnetization precession. Correspondingly, the *z*-components of the STT and SOT, $\tau_z^{\mathrm{SOT}}$ and $\tau_z^{\mathrm{STT}}$, exhibit pronounced oscillations during the pulse, reflecting their continuous dependence on the instantaneous orientation of ***m*** even though the applied current densities remain constant. However, at pulse termination, ***m*** has not yet entered the attraction basin of the reversed steady state. Upon pulse removal, ***m*** relaxes back to its initial state under the influence of the PMA field, resulting in a failed switching event. In contrast, Fig. 3(b) demonstrates that a marginal increase in pulse duration to 0.24 ns allows ***m*** to cross into the reversed state's attraction basin before the current is switched off. After the pulse ends, ***m*** relaxes deterministically toward the –*z* direction, confirming successful switching. This abrupt transition in the final state, triggered by a mere 0.01 ns extension of the pulse, underscores that ultrafast STT-SOT switching critically depends on the spin-torque-mediated precession trajectory and frequency, as illustrated in Figs. 3 (a) and 3(b).

The pulse-duration phase diagrams in Figs. 3(c) and 3(d) further corroborate this interpretation. In both parameter spaces, the minimum pulse duration required for switching decreases as current density increases, indicating that stronger spin torques enable m to traverse the dynamical separatrix in less time because of a higher precession frequency. Notably, the critical pulse duration does not scale identically with $J_{\mathrm{STT}}$ and $J_{SOT}$. As shown in Fig. 3(c), at a fixed $J_{SOT} = 5.4 \times 10^8$ A/cm$^2$, increasing $J_{STT}$ rapidly reduces the required pulse duration near the switching threshold, after which the boundary tends to saturate. In Fig. 3(d), at a fixed $J_{STT} = 3.7 \times 10^8$ A/cm$^2$, increasing $J_{SOT}$ also lowers the pulse-width threshold, but the boundary exhibits a distinct curvature. The distinct curvatures reflect the different sensitivities of the critical pulse duration to $J_{STT}$ and $J_{SOT}$, consistent with the distinct dynamical roles of the two torques discussed above. At fixed $J_{SOT}$, increasing $J_{STT}$ initially reduces the critical pulse duration substantially by enabling the SOT-excited trajectory to reach the switching condition more efficiently. As $J_{STT}$ increases further, however, this benefit gradually diminishes because the SOT-driven portion of the trajectory remains largely governed by the fixed $J_{SOT}$, flattening the boundary in Fig. 3(c). In contrast, increasing $J_{SOT}$ directly modifies the excitation and precessional evolution of $\boldsymbol{m}$, so the minimum pulse duration required to reach the switching condition continues to decrease over the entire investigated range in Fig. 3(d).

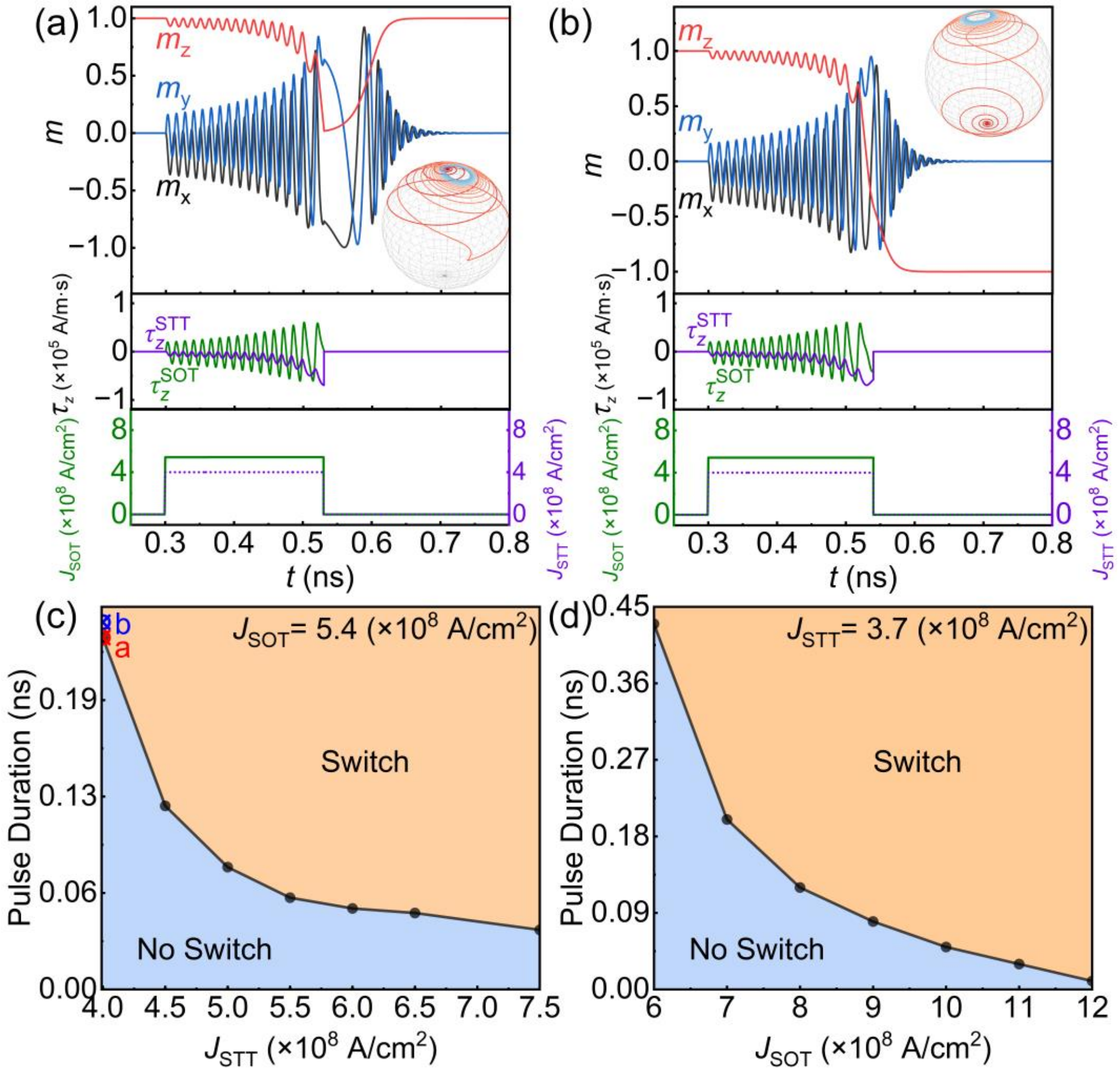

**Fig. 3 Magnetization dynamics and pulse-duration phase diagrams under combined SOT-STT driving.** (a) and (b) Time evolution of the normalized magnetization components $m_x$, $m_y$, and $m_z$ (top), the $z$-components of the effective torques $\tau_z^{\mathrm{SOT}}$ and $\tau_z^{\mathrm{STT}}$ (middle), and the applied current-density pulses $J_{\mathrm{STT}}$ and $J_{\mathrm{SOT}}$ (bottom) for two representative cases: (a) below and (b) above the switching threshold. All micromagnetic simulations use $\alpha = 0.1$ and $K_u = 1.35 \times 10^6$ J/m$^3$. Insets show the corresponding magnetization 3D trajectories on the unit sphere, colored from blue (initial state) to red (final state) to indicate time progression. (c) Switching phase diagram in the $J_{\mathrm{STT}}$-pulse-duration parameter space at fixed $J_{\mathrm{SOT}} = 5.4 \times 10^8$ A/cm$^2$. (d) Switching phase diagram in the $J_{\mathrm{SOT}}$-pulse-duration parameter space at fixed $J_{\mathrm{STT}} = 3.7 \times 10^8$ A/cm$^2$. In (c) and (d), the blue and orange shaded regions denote the no-switching and switching regimes, respectively. Marked points **a** and **b** correspond to the parameter sets used in panels (a) and (b), respectively.

## C. Switching phase diagram *vs.* material parameters

Intrinsic material properties, such as the Gilbert damping constant $\alpha$ and the uniaxial anisotropy constant $K_u$, fundamentally govern the free-layer magnetization-reversal dynamics. Moreover, extensive studies have shown that a voltage applied across the MTJ can modulate the free-layer anisotropy [36-40], underscoring the practical importance of understanding these intrinsic parameters.

Figure 4 summarizes the switching boundaries as functions of $\alpha$ and $K_u$. As shown in Fig. 4(a) and (b), the switching boundary shifts upward with increasing $\alpha$, indicating that both the critical $J_{STT}$ and the critical $J_{SOT}$ rise as damping becomes stronger. This trend is expected because stronger damping suppresses $\boldsymbol{m}$'s precessional motion, requiring larger spin torques to drive $\boldsymbol{m}$ out of its initial energy basin. Specifically, at a fixed $J_{SOT}$ = 5.4 × $10^8$ A/cm$^2$, the required $J_{STT}$ increases nearly linearly with $\alpha$. In contrast, at a fixed $J_{STT}$ = 3.7 × $10^8$ A/cm$^2$, the required $J_{SOT}$ rises rapidly at low α and then increases more gradually.

To elucidate these distinct $\alpha$-dependent trends, we derive the analytical critical conditions for the coupled STT-SOT macrospin system. We base the derivation on the LLG equation in the absence of an external magnetic field, where the critical current is determined by the loss of stability of the current-induced equilibrium state. For notational convenience, the STT and SOT effective torque-field amplitudes defined in Eq. (2) are expressed as

$$C_{\mathrm{STT}} = \frac{\hbar \eta J_{\mathrm{STT}}}{2eM_s t_F}, \quad C_{\mathrm{SOT}} = \frac{\hbar \xi J_{\mathrm{SOT}}}{2eM_s t_F} \tag{4}$$

where $\eta$ is the STT spin-polarization efficiency and $\xi$ is the damping-like SOT efficiency. The signs associated with the current polarity and spin-polarization direction are incorporated into the signed amplitudes $C_{STT}$ and $C_{SOT}$.

The magnetization direction of the free layer is written as

$$\boldsymbol{m} = (\sin\theta\cos\phi, \sin\theta\sin\phi, \cos\theta), \tag{5}$$

where $\theta$ denotes the polar angle measured from the $+z$ axis and $\phi$ the azimuthal angle in the $x$-$y$ plane. The effective perpendicular anisotropy field is given by

$$\mu_0 \boldsymbol{H}_{\mathrm{ani}} = H_k \cos\theta \hat{\boldsymbol{z}} \tag{6}$$

where $H_k = 2K_u/M_s$. In the analytical model, the perpendicular STT polarization is along the $z$ direction, and the damping-like SOT polarization is along the $y$ direction.

Substituting the angular representation of $\boldsymbol{m}$ into the LLG equation yields the angular equations of motion:

$$\begin{aligned} \dot{\theta} &= \frac{\gamma}{1+\alpha^2}\left[C_{\mathrm{STT}}\sin\theta - C_{\mathrm{SOT}}\cos\theta\sin\phi - \alpha(H_{\mathrm{k}}\sin\theta\cos\theta - C_{\mathrm{SOT}}\cos\phi)\right], \\ \dot{\phi} &= \frac{\gamma}{1+\alpha^2}\left[H_{\mathrm{k}}\cos\theta + \alpha C_{\mathrm{STT}} - C_{\mathrm{SOT}}\left(\alpha\cot\theta\sin\phi + \frac{\cos\phi}{\sin\theta}\right)\right]. \end{aligned} \tag{7}$$

At the equilibrium point $(\theta_0, \phi_0)$, $\dot{\theta} = 0$, $\dot{\phi} = 0$. Since $1+\alpha^2 > 0$, the equilibrium conditions reduce to

$$\begin{aligned} C_{\mathrm{STT}}\sin\theta_0 - C_{\mathrm{SOT}}\cos\theta_0\sin\phi_0 &= 0, \\ H_{\mathrm{k}}\sin\theta_0\cos\theta_0 - C_{\mathrm{SOT}}\cos\phi_0 &= 0. \end{aligned} \tag{8}$$

Elimination of $\phi_0$ gives

$$C_{\mathrm{SOT}}^2 = \sin^2\theta_0\left[\frac{C_{\mathrm{STT}}^2}{\cos^2\theta_0} + H_{\mathrm{k}}^2\cos^2\theta_0\right]. \tag{9}$$

To determine the stability of this equilibrium, small perturbations are introduced as

$$\theta = \theta_0 + \delta\theta, \quad \phi = \phi_0 + \delta\phi. \tag{10}$$

The linearized dynamics are

$$\frac{d}{dt}\begin{pmatrix}\delta\theta \\ \delta\phi\end{pmatrix} = \mathbf{M}\begin{pmatrix}\delta\theta \\ \delta\phi\end{pmatrix}, \tag{11}$$

where $\mathbf{M}$ is the two-dimensional Jacobian matrix of $(\dot{\theta}, \dot{\phi})$ evaluated at $(\theta_0,\phi_0)$. Its trace and determinant are

$$\mathrm{Tr}(\mathbf{M}) = \frac{\gamma}{1+\alpha^2}\left[\frac{2C_{\mathrm{STT}}}{\cos\theta_0} - \alpha H_{\mathrm{k}}\left(3\cos^2\theta_0 - 1\right)\right], \tag{12}$$

$$\det(\mathbf{M}) = \frac{\gamma^2}{1+\alpha^2}\left[\frac{C_{\mathrm{STT}}^2}{\cos^2\theta_0} + H_{\mathrm{k}}^2\cos^2\theta_0\left(2\cos^2\theta_0 - 1\right)\right]. \tag{13}$$

A locally stable equilibrium satisfies $\mathrm{Tr}(\mathbf{M}) < 0$ and $\det(\mathbf{M}) > 0$. The dynamic anti-damping instability occurs when

$$\mathrm{Tr}(\mathbf{M}) = 0, \det(\mathbf{M}) > 0, \tag{14}$$

which gives

$$C_{\mathrm{STT,c}}^{(\mathrm{d})} = \frac{\alpha H_{\mathrm{k}}}{2}\cos\theta_0(3\cos^2\theta_0 - 1), \tag{15}$$

and

$$\left|C_{\mathrm{SOT,c}}^{(\mathrm{d})}\right| = H_{\mathrm{k}} \sin\theta_0 \sqrt{\cos^2\theta_0 + \frac{\alpha^2}{4}(3\cos^2\theta_0 - 1)^2}\ . \tag{16}$$

Along this dynamic branch, the remaining condition det(**M**) > 0 restricts the equilibrium-angle parameter to $\frac{2+3\alpha^2+2\sqrt{1+\alpha^2}}{8+9\alpha^2} < \cos^2\theta_0 \le 1$.

The static instability is determined by

$$\det(\mathbf{M}) = 0, \mathrm{Tr}(\mathbf{M}) < 0\,, \tag{17}$$

and gives

$$C_{\mathrm{STT,c}}^{(\mathrm{s})} = H_{\mathrm{k}} \cos^2\theta_0 \sqrt{1-2\cos^2\theta_0}\ , \tag{18}$$

$$\left|C_{\mathrm{SOT,c}}^{(\mathrm{s})}\right| = \sqrt{2} H_{\mathrm{k}} \sin^2\theta_0 \cos\theta_0 . \tag{19}$$

Requiring a real square root in Eq. (18) and imposing the remaining condition Tr (**M**) < 0 gives the static segment, including its common endpoint with the dynamic branch, $\frac{2+3\alpha^2+2\sqrt{1+\alpha^2}}{8+9\alpha^2} \le \cos^2\theta_0 \le \frac{1}{2}$.

Equations (15)-(19) provide the parametric representation of the analytical instability boundary in terms of the equilibrium angle $\theta_0$. For comparison with the micromagnetic simulations, the equilibrium angle should be eliminated for the prescribed $J_{\mathrm{STT}}$ or $J_{\mathrm{SOT}}$. The corresponding explicit current-density expressions is given in the Appendix.

To validate the analytical model, the theoretical critical-current boundaries are superimposed on the micromagnetic simulation results, as shown by the dashed curves in Figs. 4(a) and 4(b). The analytical curves capture the main $\alpha$-dependent trends of the simulated $J_{\mathrm{STT}}$ and $J_{\mathrm{SOT}}$, supporting the interpretation that the nearly linear STT threshold is dominated by direct damping compensation, whereas the pronounced nonlinear variation of the SOT threshold arises from the coupling between the SOT amplitude, the current-induced equilibrium angle, and the stability condition. Similar comparisons are also made for the cases shown in Figs. 2(d), 4(c), and 4(d), where the analytical model consistently captures the overall trends of the micromagnetic switching boundaries. The remaining discrepancies primarily stem from the different criteria used to define the switching boundary. The analytical model predicts the theoretical boundary based on the stability of the current-induced equilibrium state, whereas the simulations employ finite current pulses and identify successful switching only when the magnetization crosses the dynamical separatrix and ultimately relaxes into the reversed-state basin. Consequently, certain current combinations slightly above the analytical boundary are still classified as non-switching in the simulations because the reversal is not completed within the pulse duration for simulations. These finite-pulse and post-pulse relaxation effects shift the simulated switching boundary above the analytical prediction without altering the overall parameter dependence.

Figures 4(c) and (d) show that the switching boundary also shifts upward with increasing $K_u$, indicating that a larger $J_{STT}$ or $J_{SOT}$ is required for switching at higher anisotropy. In both cases, the threshold current exhibits an approximately linear dependence on $K_u$. Physically, a larger $K_u$ strengthens the PMA and stabilizes the initial out-of-plane state, thereby increasing the difficulty of reversing $\boldsymbol{m}$. Consequently, stronger spin torques are required to overcome the restoring effect of the anisotropy field and drive $\boldsymbol{m}$ into the reversed stable state. Moreover, within the parameter range considered, the boundary in Fig. 4(d) varies more steeply with $K_u$ than that in Fig. 4(c), suggesting that the switching threshold of $J_{SOT}$ is more sensitive to $K_u$ than that of $J_{STT}$.

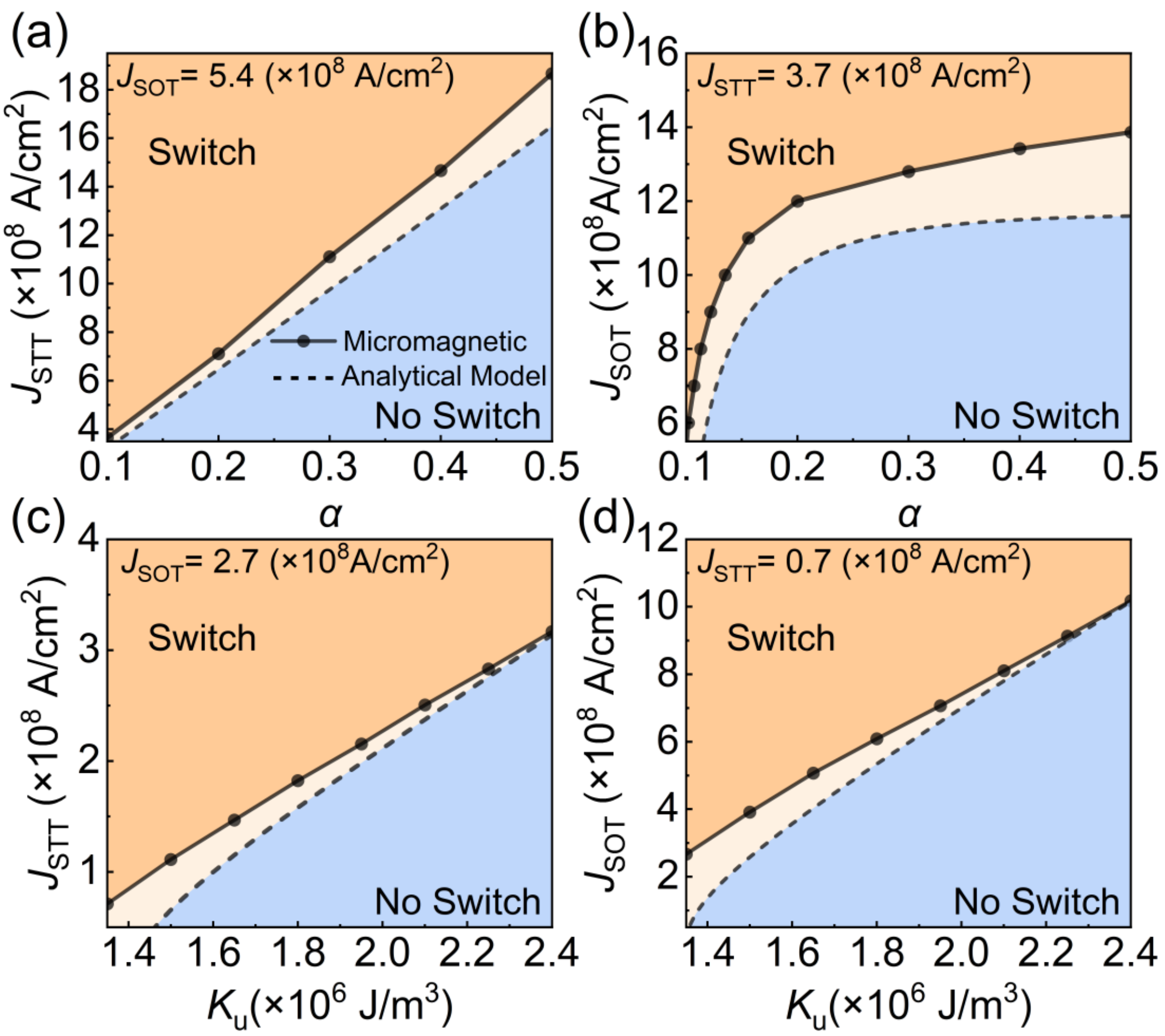

**Fig. 4 Phase diagrams delineating the switching boundaries for STT and SOT current densities as functions of material parameters.** (a) and (b) Dependence of the critical switching currents on the Gilbert damping constant $\alpha$, with the complementary current density fixed at $J_{SOT} = 5.4 \times 10^8$ A/cm$^2$ and $J_{STT} = 3.7 \times 10^8$ A/cm$^2$, respectively. (c) and (d) Dependence of the critical switching currents on the uniaxial anisotropy constant $K_u$, with the complementary current density fixed at $J_{SOT} = 2.7 \times 10^8$ A/cm$^2$ and $J_{STT} = 0.7 \times 10^8$ A/cm$^2$, respectively. In all panels, the dashed line represents the switching boundary derived from the analytical model assuming an infinite pulse duration, while the solid line with circular markers denotes the boundary obtained from micromagnetic simulations with 2 ns-duration current pulses. The blue and orange shaded regions indicate the no-switching and switching regimes, respectively; the light-shaded band between the two boundaries highlights the parameter region where the thresholds differ slightly due to the different (2 ns *vs.* infinite) pulse duration.

### D. Switching time *vs.* material parameters

The switching speed is a key metric for evaluating the write performance of MRAM devices. In an MTJ, the write operation is completed only after the free-layer magnetization reaches the target state; therefore, a shorter $\boldsymbol{m}$ reversal time directly translates to faster writing [41,42]. We further examine how material parameters and the current densities $J_{STT}$-$J_{SOT}$ influence the time required for the free layer to reach the target magnetic state. To ensure reliable switching even when the damping constant or anisotropy constant increases, we select relatively large values of $J_{STT}$ and $J_{SOT}$ in this analysis. This choice avoids scenarios where the applied current falls below the critical switching threshold after α or $K_u$ increases. The applied current densities are chosen sufficiently above the switching threshold so that changes in switching time can be studied without being confounded by switching failures near the threshold. The switching time is defined as the interval from the onset of the current pulse to the moment when the out-of-plane magnetization component reaches $m_z = -0.9$.

Figure 5(a) illustrates the magnetization dynamics for the reference case with $\alpha = 0.1$. Due to the large magnitude of the applied STT and SOT currents, the magnetization rapidly departs from the initial state (+$z$) when the current pulse is applied, unlike oscillating precession at low current density cases above [Figs. 3 and 4]. However, after entering the reversed-state basin, the $\boldsymbol{m}$ exhibits noticeable damped oscillations before fully approaching the −$z$ direction. Consequently, the switching time depends not only on when $m_z$ crosses zero but also on the time required for residual precession to decay. As shown in Fig. 5(b), increasing the damping constant keeps the overall reversal pathway similar, but strongly suppresses post-switching oscillations. Magnetization relaxes more directly toward the reversed state, and the shaded total switching-time interval becomes shorter. Therefore, although larger damping increases the critical current required for switching, it can shorten the reversal time once the applied current exceeds the threshold.

Fig. 5 (c) shows the dependence of the switching time on α. As $\alpha$ increases, the switching time decreases monotonically, but the dependence is not linear. This nonlinear behavior is expected because the damping term in the LLG equation controls the decay rate of the precessional motion. A larger $\alpha$ causes the amplitude of the residual oscillation to decay more rapidly after the magnetization has entered the reversed-state basin. The decrease in switching time with increasing $\alpha$ should therefore be understood as a post-switching relaxation effect, rather than as a reduction in the intrinsic switching threshold.

Figure 5(d) shows the influence of the $K_u$ on the switching time. When the damping constant is held constant, increasing $K_u$ also reduces the switching time within the simulated parameter range. This trend has a different physical origin than the damping dependence. A larger $K_u$ strengthens the PMA effective field, thereby increasing the restoring torque that pulls the $\boldsymbol{m}$ toward the out-of-plane easy axis. Once the current pulse drives the $\boldsymbol{m}$ into the reversed-state basin, this stronger anisotropy field accelerates its relaxation toward the $-z$ equilibrium direction. This acceleration is attributed to an increase in the characteristic precessional frequency, which shortens the time needed to complete the reversal.

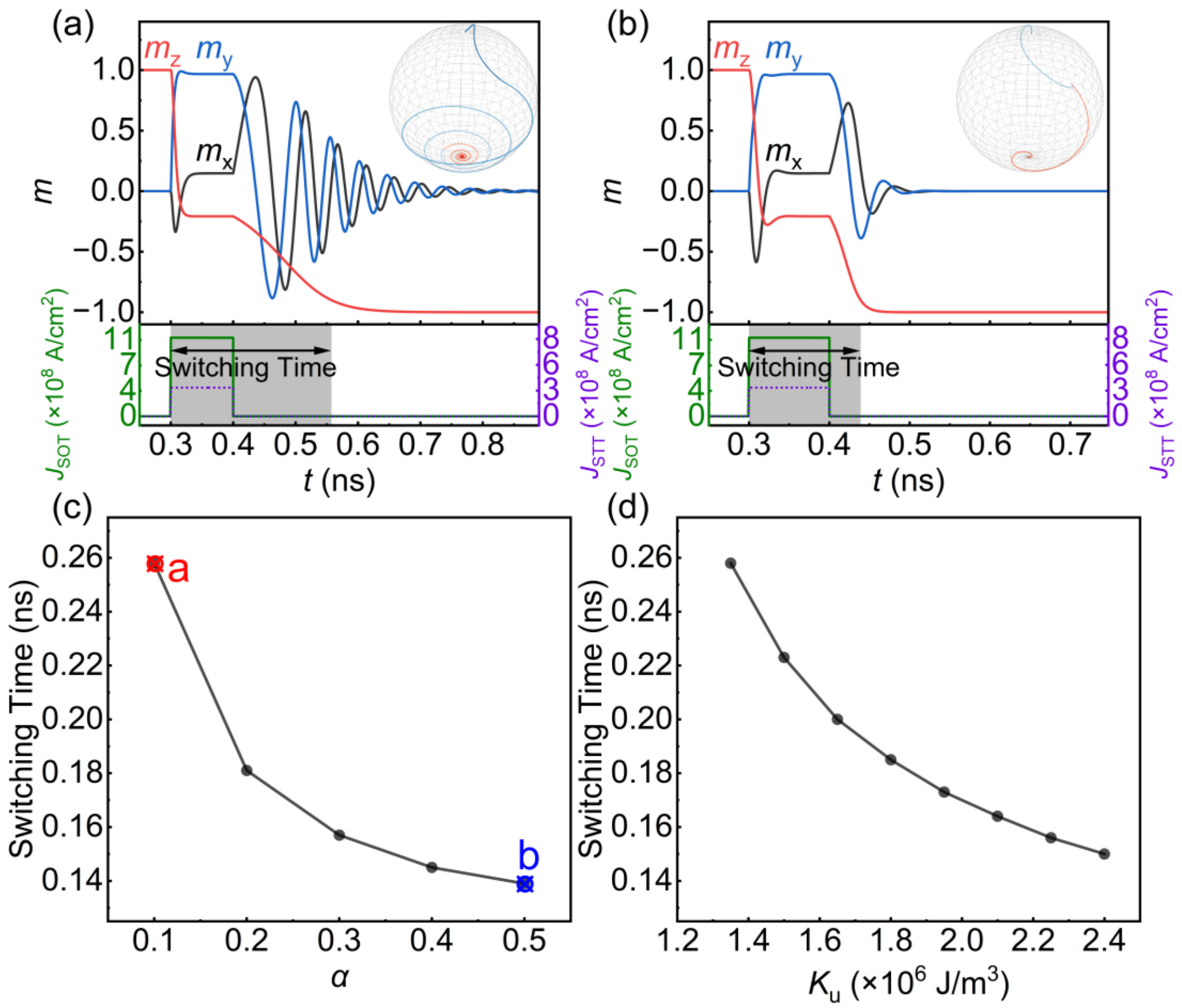

**Fig. 5 Dependence of the switching time on the Gilbert damping constant *α* and uniaxial anisotropy constant $K_u$ under combined SOT-STT driving.** (a) and (b) Temporal evolution of the magnetization components along with the applied current pulses. The shaded area marks the switching-time interval. Insets display the corresponding magnetization 3D trajectories on the unit sphere, where the color changes from blue to red as time evolves from the initial to the final state. (c) Switching time as a function of *α* with $K_u$ fixed at 1.35 × $10^6$ J/m$^3$. The points labeled **a** and **b** correspond to the specific parameter sets used for the simulations in panels (a) and (b), respectively. The current densities are fixed at $J_{STT}$ = 3.11 × $10^8$ A/cm$^2$ and $J_{SOT}$ = 11.11 × $10^8$ A/cm$^2$. (d) Switching time as a function of $K_u$, with *α* fixed at 0.1. In the above-threshold switching regime, switching time decreases as either α or $K_u$ increases.

### E. Switching time vs. current density

To further elucidate the influence of $J_{STT}$ and $J_{SOT}$ on the reversal speed, we investigated the switching dynamics under different current density combinations, with fixed $\alpha = 0.1$ and $K_u = 1.35 \times 10^6$ J/m$^3$. We compare two representative cases from the two-dimensional parameter space in Figs. 6(a) and 6(b). Figure 6(a) corresponds to a regime with relatively low $J_{STT}$ and high $J_{SOT}$. Here, the strong SOT component efficiently drives magnetization away from the initial direction, generating large-angle precession. However, by the end of the current pulse, the ***m*** remains predominantly near the in-plane region, still distant from the final perpendicular (–$z$) equilibrium. Consequently, a prolonged post-pulse relaxation period is required to complete the switch, as indicated by the broad shaded "Switching Time" interval.

In contrast, Fig. 6(b) presents the case with higher $J_{STT}$ and lower $J_{SOT}$. While the SOT-induced transverse excitation is weaker, the stronger STT component provides a more effective bias directly toward the reversed ($-z$) direction during the pulse. This results in the magnetization's trajectory reaching a state much closer to the target equilibrium by the pulse termination. As a result, the subsequent relaxation to the –$z$ state occurs more rapidly, leading to a significantly shorter overall switching time. These dynamics underscore a key finding: switching speed depends not only on how quickly the spin-torque-driven ***m*** crosses the dynamical separatrix, but also on its proximity to the final equilibrium state at pulse termination. A trajectory that finishes the pulse closer to the target state requires less post-pulse relaxation, enabling faster switching.

This interpretation is further supported by the comprehensive switching-time map in Fig. 6(c), which plots the switching time as a function of both $J_{STT}$ and $J_{SOT}$. The map reveals a clear trend: for a given $J_{SOT}$, increasing $J_{STT}$ generally reduces the switching time (transition from blue to red regions). The two specific cases from Figs. 6(a) and (b) are marked on this map, visually confirming that the configuration with higher STT contribution corresponds to faster switching. The data show that optimizing the STT-to-SOT ratio is crucial for minimizing write time in ultrafast spintronic memory operations. Therefore, within the successful-switching region of the $J_{SOT}$-$J_{STT}$ phase space, strategically increasing $J_{STT}$ while avoiding excessively large JSOT can minimize switching time, thereby enhancing the write-speed performance of the SOT/STT-MRAM.

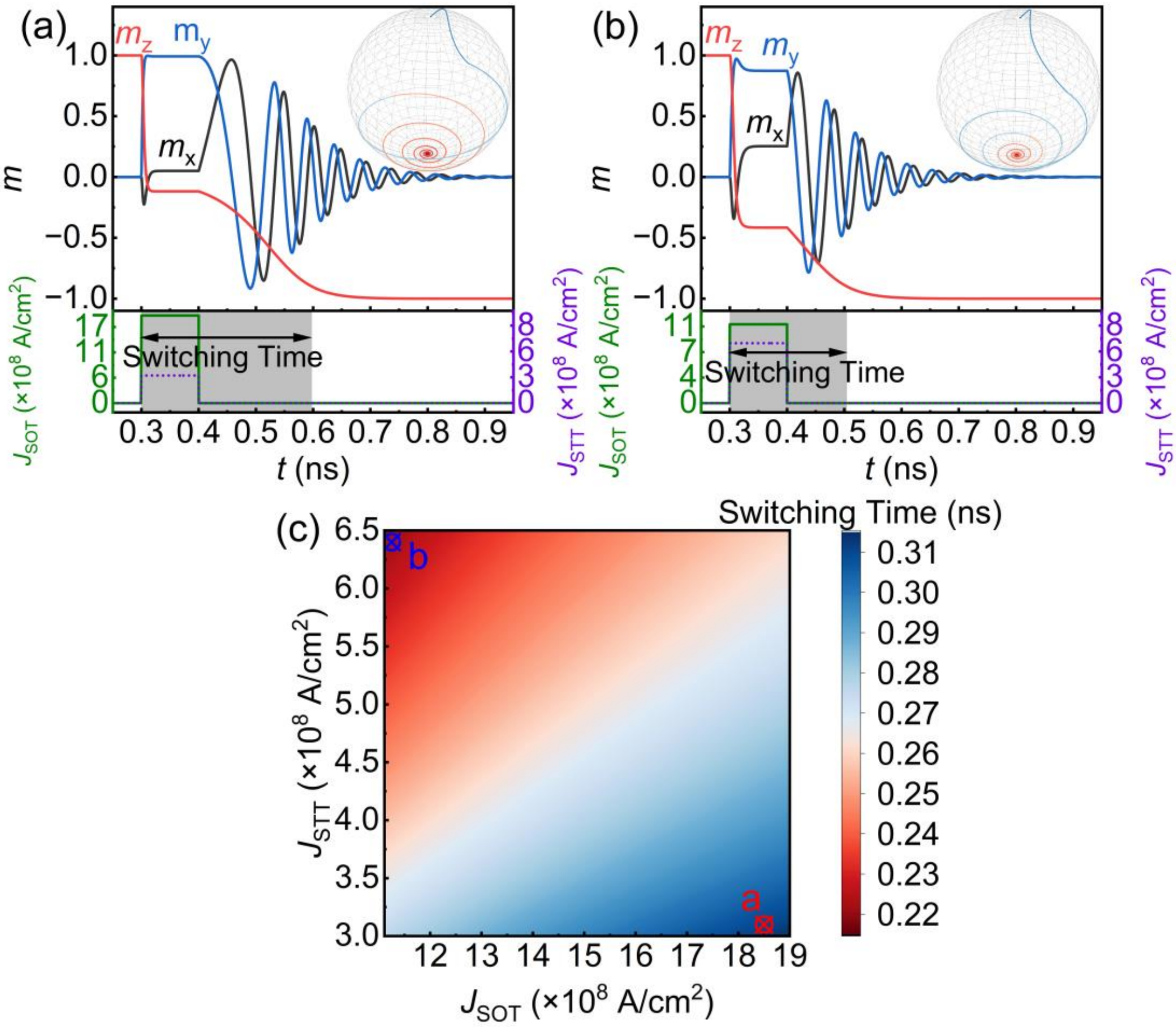

**Fig. 6 Magnetization dynamics and switching-time distribution in the $J_{SOT}$-$J_{STT}$ parameter space.** (a) and (b) Temporal evolution of the magnetization components under the applied SOT and STT current pulses for two representative operating points: (a) $J_{STT} = 3 \times 10^8$ A cm$^{-2}$ and $J_{SOT} = 19 \times 10^8$ A cm$^{-2}$; (b) $J_{STT} = 6.5 \times 10^8$ A cm$^{-2}$ and $J_{SOT} = 11.11 \times 10^8$ A cm$^{-2}$. The shaded region marks the switching-time interval. The insets display the corresponding magnetization 3D trajectories on the unit sphere, with the trace color evolving from blue to red as time progresses from the initial to the final state. (c) Switching-time map as a function of $J_{SOT}$ and $J_{STT}$. The color scale indicates the switching time, with points a and b corresponding to the cases shown in panels (a) and (b), respectively.

## IV. Conclusion

In summary, we systematically investigated a field-free ultrafast magnetization-switching scheme that combines SOT and STT in a perpendicular magnetic tunnel junction. The two torques play complementary but non-equivalent dynamical roles: SOT efficiently drives the magnetization away from its initial perpendicular state and promotes large-angle precessional motion, whereas STT provides a perpendicular biasing torque that steers the evolving trajectory toward the reversed state and facilitates final-state selection. Their cooperative action yields a distinctly nonlinear switching boundary in the $J_{STT}$–$J_{SOT}$ parameter space, showing that reversal is governed by trajectory evolution rather than a simple additive superposition of torque amplitudes. An analytical macrospin model, based on the stability analysis of the current-induced equilibrium, captures the overall switching-boundary trends obtained from micromagnetic simulations across a broad range of current densities, damping constants, and anisotropy strengths, and distinguishes dynamic anti-damping and static instability branches. This provides a unified theoretical description of the critical conditions for cooperative STT–SOT switching.

Our analysis further reveals that successful, fast reversal depends not only on torque amplitude but also critically on pulse duration and, most importantly, the magnetization state at pulse termination. Successful switching requires the magnetization to enter the reversed-state basin before the pulse ends, while a pulse-end state positioned closer to the final equilibrium minimizes subsequent relaxation and enables faster reversal. The systematic dependencies on damping, magnetic anisotropy, and the STT–SOT balance underscore that the switching threshold and post-pulse dynamics must be considered jointly. The STT–SOT balance offers an additional degree of freedom for tailoring the pulse-end trajectory: an appropriately weighted STT contribution can steer the pulse-end state toward the target direction, whereas merely increasing SOT does not necessarily accelerate switching.

These findings establish that the pulse-end magnetization state, rather than the instantaneous torque amplitude, is the decisive factor governing switching speed and reliability in coupled STT–SOT systems. The cooperative mechanism demonstrated here provides a physical framework for understanding and designing field-free switching in perpendicular magnetic heterostructures, and offers practical guidance for optimizing three-terminal spintronic devices.

**Acknowledgements**

We acknowledge support from the National Key Research and Development Program of China (Grant No. 2023YFA1406603, 2024YFA1408801), the National Natural Science Foundation of China (Grant No. 12574125).

**Appendix: explicit critical-current expressions and parameter values substitution**

The main text provides the static and dynamic instability conditions and the corresponding ranges of the equilibrium-angle parameter. Here, we eliminate $\theta_0$ to obtain the critical current density for a prescribed STT or SOT current. A superscript asterisk denotes a prescribed current density.

**A. Fixed $J_{\mathrm{STT}}$**

For a prescribed STT current density $J_{\mathrm{STT}} = J^*_{\mathrm{STT}}$, the corresponding torque-field amplitude is

$$C^*_{\mathrm{STT}} = \frac{\hbar \eta J^*_{\mathrm{STT}}}{2eM_{\mathrm{s}}t_{\mathrm{F}}}. \quad \text{(A1)}$$

**1. Static branch**

For the static branch, Eq. (18) then becomes

$$C^*_{\mathrm{STT}} = H_{\mathrm{k}} \cos^2\theta_0 \sqrt{1-2\cos^2\theta_0}\,. \quad \text{(A2)}$$

Squaring both sides gives

$$2\cos^6\theta_0 - \cos^4\theta_0 + \left(\frac{C^*_{\mathrm{STT}}}{H_{\mathrm{k}}}\right)^2 = 0\,. \quad \text{(A3)}$$

In the pure-SOT limit $J^*_{\mathrm{STT}} = 0$, and hence from Eq. (4). Within the static-branch range specified in the main text, Eq. (A3) then gives the relevant endpoint $\cos^2\theta_0 = 1/2$. Therefore, introducing $a \equiv \cos^2\theta_0$, we select the root that continuously approaches $a = 1/2$ as $J^*_{\mathrm{STT}} \to 0$:

$$a \equiv \cos^2\theta_0 = \frac{1}{6} + \frac{1}{3}\cos\left\{\frac{1}{3}\arccos\left[1-54\left(\frac{\hbar\eta J^*_{\mathrm{STT}}}{2eM_{\mathrm{s}}t_{\mathrm{F}}H_{\mathrm{k}}}\right)^2\right]\right\}. \quad \text{(A4)}$$

Substitution into Eq. (19) then gives the corresponding critical SOT current density,

$$\left|J_{\mathrm{SOT,c}}^{(\mathrm{s})}\right| = \frac{2eM_{\mathrm{s}}t_{\mathrm{F}}}{\hbar|\xi|}\sqrt{2}H_{\mathrm{k}}\sqrt{a}(1-a)\,. \tag{A5}$$

The root in Eq. (A4) is retained only within the static-branch range specified after Eq. (19) in the main text.

**2. Dynamic branch**

For the dynamic branch, substituting the prescribed $C_{\mathrm{STT}}^{*}$ into Eq. (15) gives

$$\frac{\hbar\eta J_{\mathrm{STT}}^{*}}{2eM_{\mathrm{s}}t_{\mathrm{F}}} = \frac{\alpha H_{\mathrm{k}}}{2}\cos\theta_0\left(3\cos^2\theta_0 - 1\right). \tag{A6}$$

Rearranging gives

$$3\cos^3\theta_0 - \cos\theta_0 = \frac{\hbar\eta J_{\mathrm{STT}}^{*}}{eM_{\mathrm{s}}t_{\mathrm{F}}\alpha H_{\mathrm{k}}}. \tag{A7}$$

At the pure-STT endpoint of the dynamic branch, $J_{\mathrm{SOT}} = 0$, and hence $C_{\mathrm{SOT}} = 0$ from Eq. (4). Within the dynamic-branch range specified in the main text, Eq. (16) then gives the relevant endpoint $\cos\theta_0 = 1$. Therefore, introducing $b \equiv \cos\theta_0$, we select the root of Eq. (A7) that continuously approaches b = 1 as $J_{\mathrm{SOT}}^{*} \to 0$. This root can be written as

$$\begin{aligned} b \equiv \cos\theta_0 = & \sqrt[3]{\frac{\hbar\eta J_{\mathrm{STT}}^{*}}{6eM_{\mathrm{s}}t_{\mathrm{F}}\alpha H_{\mathrm{k}}} + \sqrt{\left(\frac{\hbar\eta J_{\mathrm{STT}}^{*}}{6eM_{\mathrm{s}}t_{\mathrm{F}}\alpha H_{\mathrm{k}}}\right)^2 - \frac{1}{729}}} \\ & + \sqrt[3]{\frac{\hbar\eta J_{\mathrm{STT}}^{*}}{6eM_{\mathrm{s}}t_{\mathrm{F}}\alpha H_{\mathrm{k}}} - \sqrt{\left(\frac{\hbar\eta J_{\mathrm{STT}}^{*}}{6eM_{\mathrm{s}}t_{\mathrm{F}}\alpha H_{\mathrm{k}}}\right)^2 - \frac{1}{729}}}\,. \end{aligned} \tag{A8}$$

The selected root must also satisfy the dynamic-branch range specified after Eq. (16). Substitution of $b$ into Eq. (16) gives

$$\left|J_{\mathrm{SOT,c}}^{(\mathrm{d})}\right| = \frac{2eM_{\mathrm{s}}t_{\mathrm{F}}H_{\mathrm{k}}}{\hbar|\xi|}\sqrt{1-b^2}\sqrt{b^2 + \frac{\alpha^2}{4}(3b^2-1)^2}\,. \tag{A9}$$

At the pure-STT endpoint, $J_{\mathrm{SOT}} = 0$ and $b = 1$, as obtained above. Substituting b = 1 into Eq. (15) gives $C_{\mathrm{STT,c}}^{\mathrm{pure}} = \alpha H_{\mathrm{k}}$. If the prescribed STT current reaches the pure-STT instability threshold,

$$J_{\mathrm{STT}}^{*} \geq J_{\mathrm{STT,c}}^{\mathrm{pure}} = \frac{2eM_{\mathrm{s}}t_{\mathrm{F}}}{\hbar\eta}\alpha H_{\mathrm{k}}, \tag{A10}$$

the STT alone is sufficient to compensate the Gilbert damping and destabilize the equilibrium connected to the initial $+\hat{\boldsymbol{z}}$ state. In this regime, no additional SOT is required to trigger the instability $y$, and therefore $\left|J_{\mathrm{SOT,c}}\right| = 0$.

Combining the static and dynamic branches, the critical SOT current density for a prescribed $J^*_{\mathrm{STT}}$ can be written in the following piecewise form:

$$\left|J_{\mathrm{SOT,c}}\right| = \begin{cases} \dfrac{2eM_{\mathrm{s}}t_{\mathrm{F}}H_{\mathrm{k}}}{\hbar|\xi|}\sqrt{2}(1-a)\sqrt{a}, & 0 \le J^*_{\mathrm{STT}} \le \dfrac{eM_{\mathrm{s}}t_{\mathrm{F}}\alpha H_{\mathrm{k}}}{\hbar\eta}\sqrt{r}(3r-1), \\ \dfrac{2eM_{\mathrm{s}}t_{\mathrm{F}}H_{\mathrm{k}}}{\hbar|\xi|}\sqrt{1-b^2}\sqrt{b^2+\dfrac{\alpha^2}{4}(3b^2-1)^2}, & \dfrac{eM_{\mathrm{s}}t_{\mathrm{F}}\alpha H_{\mathrm{k}}}{\hbar\eta}\sqrt{r}(3r-1) < J^*_{\mathrm{STT}} < \dfrac{2eM_{\mathrm{s}}t_{\mathrm{F}}\alpha H_{\mathrm{k}}}{\hbar\eta}, \\ 0, & J^*_{\mathrm{STT}} \ge \dfrac{2eM_{\mathrm{s}}t_{\mathrm{F}}\alpha H_{\mathrm{k}}}{\hbar\eta}, \end{cases} \quad \text{(A11)}$$

where $r = \dfrac{2+3\alpha^2+2\sqrt{1+\alpha^2}}{8+9\alpha^2}$.

**B. Fixed $J_{\mathrm{SOT}}$**

For a prescribed SOT current density $J_{\mathrm{SOT}} = J^*_{\mathrm{SOT}}$, the corresponding torque-field amplitude is

$$\left|C^*_{\mathrm{SOT}}\right| = \frac{\hbar\left|\xi J^*_{\mathrm{SOT}}\right|}{2eM_{\mathrm{s}}t_{\mathrm{F}}}. \quad \text{(A12)}$$

At the pure-SOT limit, $J_{\mathrm{STT}} = 0$, and hence $C_{\mathrm{STT}} = 0$ from Eq. (4). Within the static-branch range specified in the main text, Eq. (18) then gives $\cos^2\theta_0 = 1/2$. Substituting this value into Eq. (19) gives $C^{\mathrm{pure}}_{\mathrm{SOT,c}} = \dfrac{H_{\mathrm{k}}}{2}$. Using Eq. (4), the corresponding pure-SOT critical current density is therefore $\left|J^{\mathrm{pure}}_{\mathrm{SOT,c}}\right| = \dfrac{eM_{\mathrm{s}}t_{\mathrm{F}}H_{\mathrm{k}}}{\hbar|\xi|}$.

For $0 < \left|J^*_{\mathrm{SOT}}\right| \le \dfrac{eM_{\mathrm{s}}t_{\mathrm{F}}H_{\mathrm{k}}}{\hbar|\xi|}$, SOT alone does not reach the instability threshold, and the minimum additional STT is determined by the dynamic branch. Substituting the prescribed $C^*_{\mathrm{SOT}}$ into Eq. (16) gives

$$\left|C^*_{\mathrm{SOT}}\right| = H_{\mathrm{k}}\sin\theta_0\sqrt{\cos^2\theta_0+\frac{\alpha^2}{4}(3\cos^2\theta_0-1)^2}. \quad \text{(A13)}$$

Squaring both sides and using $\sin^2\theta_0 = 1-\cos^2\theta_0$ gives

$$\left(\frac{\left|C^*_{\mathrm{SOT}}\right|}{H_{\mathrm{k}}}\right)^2 = (1-\cos^2\theta_0)\left[\cos^2\theta_0+\frac{\alpha^2}{4}(3\cos^2\theta_0-1)^2\right]. \quad \text{(A14)}$$

Introducing $c \equiv \cos^2\theta_0$ and expanding Eq. (A14) gives

$$9\alpha^2c^3+(4-15\alpha^2)c^2+(7\alpha^2-4)c+4\left(\frac{\hbar\left|\xi J_{\mathrm{SOT}}^*\right|}{2eM_\mathrm{s}t_\mathrm{F}H_\mathrm{k}}\right)^2-\alpha^2=0. \tag{A15}$$

At the pure-STT endpoint, $J_{\mathrm{SOT}}^*=0$, and hence $C_{\mathrm{SOT}}^*=0$ from Eq. (4). Within the dynamic-branch range specified in the main text, Eq. (A14) then gives the relevant endpoint $c=\cos^2\theta_0=1$. Therefore, we select the real root of Eq. (A16) that continuously approaches $c=1$ as $J_{\mathrm{SOT}}^*\to 0$.

For an explicit expression of this root, we define

$$\Delta_0=16-12\alpha^2+36\alpha^4, \tag{A16}$$

and

$$\Delta_1=128-144\alpha^2-1728\alpha^4-432\alpha^6+8748\alpha^4\left(\frac{\hbar\,|\,\xi\,|\,J_{\mathrm{SOT}}^*}{2eM_st_FH_k}\right)^2. \tag{A17}$$

The selected root can then be written as

$$c=-\frac{1}{27\alpha^2}\left[4-15\alpha^2+2\sqrt{\Delta_0}\cos\left(\frac{1}{3}\arccos\frac{\Delta_1}{2\Delta_0^{3/2}}+\frac{2\pi}{3}\right)\right]. \tag{A18}$$

The root in Eq. (A18) is taken within the dynamic-branch range specified after Eq. (16) and is continuously connected to $c=1$. For the minimum-STT solution below the pure-SOT threshold considered here, this corresponds to $\frac{1}{2}<c\leq 1$.

Substituting c = $\cos^2\theta_0$ into Eq. (15) and converting the torque-field amplitude using Eq. (4) gives the dynamic STT critical current density,

$$J_{\mathrm{STT,c}}^{(\mathrm{d})}=\frac{eM_\mathrm{s}t_\mathrm{F}\alpha H_\mathrm{k}}{\hbar\eta}\sqrt{c}(3c-1) \tag{A19}$$

Once $J_{\mathrm{SOT}}^*$ reaches the pure-SOT threshold, SOT alone reaches the instability threshold, and no additional STT is required. Therefore, the minimum additional STT required for a prescribed $J_{\mathrm{SOT}}^*$ is

$$J_{\mathrm{STT,c}}=\begin{cases}\dfrac{eM_\mathrm{s}t_\mathrm{F}\alpha H_\mathrm{k}}{\hbar\eta}\sqrt{c}(3c-1), & 0\leq\left|J_{\mathrm{SOT}}^*\right|<\dfrac{eM_\mathrm{s}t_\mathrm{F}H_\mathrm{k}}{\hbar\left|\xi\right|},\\[2ex] 0, & \left|J_{\mathrm{SOT}}^*\right|\geq\dfrac{eM_\mathrm{s}t_\mathrm{F}H_\mathrm{k}}{\hbar\left|\xi\right|}.\end{cases} \tag{A20}$$

**Author Declarations**

## Conflict of Interest

The authors have no conflicts to disclose.

## Data Availability

The data that support the findings of this study are available from the corresponding author upon reasonable request.

## References

[1] A. Manchon, J. Železný, I. M. Miron, T. Jungwirth, J. Sinova, A. Thiaville, K. Garello, and P. Gambardella, Current-induced spin-orbit torques in ferromagnetic and antiferromagnetic systems, Rev. Mod. Phys. **91**, 035004 (2019). https://doi.org/10.1103/RevModPhys.91.035004.

[2] B. Dieny, I. L. Prejbeanu, K. Garello, P. Gambardella, P. Freitas, R. Lehndorff, W. Raberg, U. Ebels, S. O. Demokritov, J. Akerman, *et al.*, Opportunities and challenges for spintronics in the microelectronics industry, Nat. Electron. **3**, 446 (2020). https://doi.org/10.1038/s41928-020-0461-5.

[3] H. Yang, S. O. Valenzuela, M. Chshiev, S. Couet, B. Dieny, B. Dlubak, A. Ferk, K. Garello, M. Jamet, D.-E. Jeong, *et al.*, Two-dimensional materials prospects for non-volatile spintronic memories, Nature **606**, 663 (2022). https://doi.org/10.1038/s41586-022-04768-0.

[4] D. C. Worledge and G. Hu, Spin-transfer torque magnetoresistive random access memory technology status and future directions, Nat. Rev. Electr. Eng. **1**, 730 (2024). https://doi.org/10.1038/s44287-024-00111-z.

[5] L. Liu, T. Moriyama, D. C. Ralph, and R. A. Buhrman, Spin-Torque Ferromagnetic Resonance Induced by the Spin Hall Effect, Phys. Rev. Lett. **106**, 036601 (2011). https://doi.org/10.1103/PhysRevLett.106.036601.

[6] L. Liu, C.-F. Pai, Y. Li, H. W. Tseng, D. C. Ralph, and R. A. Buhrman, Spin-Torque Switching with the Giant Spin Hall Effect of Tantalum, Science **336**, 555 (2012). https://doi.org/10.1126/science.1218197.

[7] M. Cubukcu, O. Boulle, M. Drouard, K. Garello, C. Onur Avci, I. Mihai Miron, J. Langer, B. Ocker, P. Gambardella, and G. Gaudin, Spin-orbit torque magnetization switching of a three-terminal perpendicular magnetic tunnel junction, Appl. Phys. Lett. **104**, 042406 (2014). https://doi.org/10.1063/1.4863407.

[8] L. Yang, Y. Fei, K. Zhou, L. Chen, Q. Fu, L. Li, C. Yan, H. Li, Y. Du, and R. Liu, Maximizing spin–orbit torque efficiency of Ta(O)/Py via modulating oxygen-induced interface orbital hybridization, Appl. Phys. Lett. **118**, 032405 (2021). https://doi.org/10.1063/5.0033752.

[9] Q. Fu, L. Liang, W. Wang, L. Yang, K. Zhou, Z. Li, C. Yan, L. Li, H. Li, and R. Liu, Observation of nontrivial spin-orbit torque in single-layer ferromagnetic metals, Phys. Rev. B **105**, 224417 (2022). https://doi.org/10.1103/PhysRevB.105.224417.

[10] Z. Tao, Z. Zhu, H. Duan, H. Wang, Z. Li, M. Yin, L. Chen, L. Ni, B. Liu, T. Zhou, Y. Xu, and R. Liu, Achieving high spin–orbit torque efficiency and robust thermal stability in the sputter-grown topological bilayer BiSb/Pt, Appl. Phys. Lett. **128**, 212409 (2026). https://doi.org/10.1063/5.0323975.
[11] S. Fukami, C. Zhang, S. DuttaGupta, A. Kurenkov, and H. Ohno, Magnetization switching by spin–orbit torque in an antiferromagnet–ferromagnet bilayer system, Nat. Mater. **15**, 535 (2016). https://doi.org/10.1038/nmat4566.
[12] Y. C. Wu, K. Garello, W. Kim, M. Gupta, M. Perumkunnil, V. Kateel, S. Couet, R. Carpenter, S. Rao, S. Van Beek, *et al.*, Voltage-Gate-Assisted Spin-Orbit-Torque Magnetic Random-Access Memory for High-Density and Low-Power Embedded Applications, Phys. Rev. Appl. **15**, 064015 (2021). https://doi.org/10.1103/PhysRevApplied.15.064015.
[13] K. Cai, G. Talmelli, K. Fan, S. Van Beek, V. Kateel, M. Gupta, M.G. Monteiro, M. Ben Chroud, G. Jayakumar, A. Trovato, *et al.*, *First Demonstration of Field-Free Perpendicular SOT-MRAM for Ultrafast and High-Density Embedded Memories*, in *2022 International Electron Devices Meeting (IEDM)* (2022), pp. 36.2.1-36.2.4. https://doi.org/10.1109/IEDM45625.2022.10019360.
[14] V. Krizakova, M. Perumkunnil, S. Couet, P. Gambardella, and K. Garello, Spin-orbit torque switching of magnetic tunnel junctions for memory applications, J. Magn. Magn. Mater. **562**, 169692 (2022). https://doi.org/10.1016/j.jmmm.2022.169692.
[15] L. Liu, O. J. Lee, T. J. Gudmundsen, D. C. Ralph, and R. A. Buhrman, Current-Induced Switching of Perpendicularly Magnetized Magnetic Layers Using Spin Torque from the Spin Hall Effect, Phys. Rev. Lett. **109**, 096602 (2012). https://doi.org/10.1103/PhysRevLett.109.096602.
[16] L. Liu, C. Zhou, X. Shu, C. Li, T. Zhao, W. Lin, J. Deng, Q. Xie, S. Chen, J. Zhou, *et al.*, Symmetry-dependent field-free switching of perpendicular magnetization, Nat. Nanotechnol. **16**, 277 (2021). https://doi.org/10.1038/s41565-020-00826-8.
[17] I.-H. Kao, R. Muzzio, H. Zhang, M. Zhu, J. Gobbo, S. Yuan, D. Weber, R. Rao, J. Li, J. H. Edgar, *et al.*, Deterministic switching of a perpendicularly polarized magnet using unconventional spin–orbit torques in $WTe_2$, Nat. Mater. **21**, 1029 (2022). https://doi.org/10.1038/s41563-022-01275-5.
[18] H. Duan, H. Gu, Z. Zhu, Z. Li, Y. Huang, L. Chen, M. Yin, L. Ni, K. Xu, X. Wang, Y. Xu, B. Liu, T. Zhou, and R. Liu, Copper-doped sputtered BiSb topological insulators for field-free perpendicular magnetization switching via out-of-plane spin polarization, Phys. Rev. Mater. **10**, 064201 (2026). https://doi.org/10.1103/s1v5-qrtc.
[19] Z. Zheng, Y. Zhang, V. Lopez-Dominguez, L. Sánchez-Tejerina, J. Shi, X. Feng, L. Chen, Z. Wang, Z. Zhang, K. Zhang, *et al.*, Field-free spin-orbit torque-induced switching of perpendicular magnetization in a ferrimagnetic layer with a vertical composition gradient, Nat. Commun. **12**, 4555 (2021). https://doi.org/10.1038/s41467-021-24854-7.
[20] X. Shu, L. Liu, J. Zhou, W. Lin, Q. Xie, T. Zhao, C. Zhou, S. Chen, H. Wang, J. Chai, *et al.*, Field-Free Switching of Perpendicular Magnetization Induced by Longitudinal Spin-Orbit-Torque Gradient, Phys. Rev. Appl. **17**, 024031 (2022). https://doi.org/10.1103/PhysRevApplied.17.024031.

[21] V. D. Nguyen, S. Rao, K. Wostyn, and S. Couet, Recent progress in spin-orbit torque magnetic random-access memory, npj Spintronics **2**, 48 (2024). https://doi.org/10.1038/s44306-024-00044-1.
[22] S. M. Vlasov, G. J. Kwiatkowski, I. S. Lobanov, V. M. Uzdin, and P. F. Bessarab, Optimal protocol for spin-orbit torque switching of a perpendicular nanomagnet, Phys. Rev. B **105**, 134404 (2022). https://doi.org/10.1103/PhysRevB.105.134404.
[23] T. Taniguchi, Suppression of writing error in a spin-orbit magnetization switching device using fieldlike torque, Phys. Rev. B **112**, 014411 (2025). https://doi.org/10.1103/yngx-z13v.
[24] C. Abert, H. Sepehri-Amin, F. Bruckner, C. Vogler, M. Hayashi, and D. Suess, Back-Hopping in Spin-Transfer-Torque Devices: Possible Origin and Countermeasures, Phys. Rev. Appl. **9**, 054010 (2018). https://doi.org/10.1103/PhysRevApplied.9.054010.
[25] M. Hoffmann, V. Krizakova, V. Kateel, K. Cai, S. Couet, and P. Gambardella, Pulse-shaping strategies for efficient switching of magnetic tunnel junctions by spin-orbit torque, Phys. Rev. Appl. **22**, 034052 (2024). https://doi.org/10.1103/PhysRevApplied.22.034052.
[26] K. Ray, J. Vigier, P. Usé, S. Martin, N. Lefoulon, C. Bouard, M. Drouard, and G. Gaudin, Intrinsic back-switching phenomenon in spin-orbit torque MRAM devices, Phys. Rev. Appl. **24**, 064038 (2025). https://doi.org/10.1103/vlhq-tf18.
[27] M. Wang, W. Cai, D. Zhu, Z. Wang, J. Kan, Z. Zhao, K. Cao, Z. Wang, Y. Zhang, T. Zhang, C. Park, J. Wang, A. Fert, and W. Zhao, Field-free switching of a perpendicular magnetic tunnel junction through the interplay of spin–orbit and spin-transfer torques, Nat. Electron. **1**, 582 (2018). https://doi.org/10.1038/s41928-018-0160-7.
[28] E. Grimaldi, V. Krizakova, G. Sala, F. Yasin, S. Couet, G. Sankar Kar, K. Garello, and P. Gambardella, Single-shot dynamics of spin–orbit torque and spin transfer torque switching in three-terminal magnetic tunnel junctions, Nat. Nanotechnol. **15**, 111 (2020). https://doi.org/10.1038/s41565-019-0607-7.
[29] A. Meo, J. Chureemart, R. W. Chantrell, and P. Chureemart, Magnetisation switching dynamics induced by combination of spin transfer torque and spin orbit torque, Sci. Rep. **12**, 3380 (2022). https://doi.org/10.1038/s41598-022-07277-2.
[30] J. C. Slonczewski, Current-driven excitation of magnetic multilayers, J. Magn. Magn. Mater. **159**, L1 (1996). https://doi.org/10.1016/0304-8853(96)00062-5.
[31] M. J. Donahue and D. G. Porter, *OOMMF User's Guide, Version 1.0*, NISTIR 6376 (National Institute of Standards and Technology, Gaithersburg, MD, 1999). https://doi.org/10.6028/NIST.IR.6376.
[32] T. L. Gilbert, A phenomenological theory of damping in ferromagnetic materials, IEEE Trans. Magn. **40**, 3443 (2004). https://doi.org/10.1109/TMAG.2004.836740.
[33] A. Aharoni, *Introduction to the Theory of Ferromagnetism*, 2nd Ed. (Oxford University Press, Oxford, 2001).
[34] D. Bedau, H. Liu, J. Z. Sun, J. A. Katine, E. E. Fullerton, S. Mangin, and A. D. Kent, Spin-transfer pulse switching: From the dynamic to the thermally activated regime, Appl. Phys. Lett. **97**, 262502 (2010). https://doi.org/10.1063/1.3532960.
[35] D. Apalkov, B. Dieny, and J. M. Slaughter, Magnetoresistive Random Access

Memory, Proc. IEEE **104**, 1796 (2016). https://doi.org/10.1109/JPROC.2016.2590142.
[36] Y. Shiota, T. Nozaki, F. Bonell, S. Murakami, T. Shinjo, and Y. Suzuki, Induction of coherent magnetization switching in a few atomic layers of FeCo using voltage pulses, Nat. Mater. **11**, 39 (2012). https://doi.org/10.1038/nmat3172.
[37] V. Krizakova, E. Grimaldi, K. Garello, G. Sala, S. Couet, G. S. Kar, and P. Gambardella, Interplay of Voltage Control of Magnetic Anisotropy, Spin-Transfer Torque, and Heat in the Spin-Orbit-Torque Switching of Three-Terminal Magnetic Tunnel Junctions, Phys. Rev. Appl. **15**, 054055 (2021). https://doi.org/10.1103/PhysRevApplied.15.054055.
[38] T. Zhang, C. Wan, and X. Han, Threshold current of field-free perpendicular magnetization switching using anomalous spin-orbit torque, Phys. Rev. B **108**, 014432 (2023). https://doi.org/10.1103/PhysRevB.108.014432.
[39] T. Yamamoto, T. Ichinose, T. Nozaki, S. Tamaru, K. Yakushiji, H. Kubota, and S. Yuasa, Write-error reduction in voltage-driven magnetization switching using a recording layer with low magnetic damping, Phys. Rev. Appl. **21**, 054008 (2024). https://doi.org/10.1103/PhysRevApplied.21.054008.
[40] D. Favaro et al., Switching of a magnetic tunnel junction by voltage-controlled magnetic anisotropy, Phys. Rev. Appl. **23**, 014040 (2025). https://doi.org/10.1103/PhysRevApplied.23.014040.
[41] K. Garello, C. O. Avci, I. M. Miron, M. Baumgartner, A. Ghosh, S. Auffret, O. Boulle, G. Gaudin, and P. Gambardella, Ultrafast magnetization switching by spin-orbit torques, Appl. Phys. Lett. **105**, 212402 (2014). https://doi.org/10.1063/1.4902443.
[42] Q. Yang, D. Han, S. Zhao, J. Kang, F. Wang, S.-C. Lee, J. Lei, K.-J. Lee, B.-G. Park, and H. Yang, Field-free spin–orbit torque switching in ferromagnetic trilayers at sub-ns timescales, Nat. Commun. **15**, 1814 (2024). https://doi.org/10.1038/s41467-024-46113-1.